\documentclass[fleqn,twoside]{article}
\usepackage{myespcrc2}

\usepackage{graphicx}
\usepackage{hyperref}
\usepackage[figuresright]{rotating}

\newcommand{\AmS}{{\protect\the\textfont2
  A\kern-.1667em\lower.5ex\hbox{M}\kern-.125emS}}
  
 \newcommand{\nonubb}  {$0 \nu \beta \beta$}

\newcommand{\bb}{$\beta\beta$}

\newcommand{\MJ}{{\sc{Majo\-ra\-na}}}
\def\DEM{{\sc Demonstrator}}

\def\ppc{P-PC}  
\def\nuc#1#2{${}^{#1}$#2}

\title{The {\sc Majorana} Experiment}

\newcommand{\alberta}{Centre for Particle Physics, University of Alberta, Edmonton, AB, Canada}
\newcommand{\blhill}{Department of Physics, Black Hills State University, Spearfish, SD, USA}
\newcommand{\ITEP}{Institute for Theoretical and Experimental Physics, Moscow, Russia}
\newcommand{\JINR}{Joint Institute for Nuclear Research, Dubna, Russia}
\newcommand{\lbnl}{Lawrence Berkeley National Laboratory, Berkeley, CA, USA}
\newcommand{\lanl}{Los Alamos National Laboratory, Los Alamos, NM, USA}

\newcommand{\uw}{Center for Experimental Nuclear Physics and Astrophysics, 
and Dept. of Physics, University of Washington, Seattle, WA, USA}
\newcommand{\uchic}{Department of Physics, University of Chicago, Chicago, IL, USA}
\newcommand{\unc}{Department of Physics, University of North Carolina, Chapel Hill, NC, USA}
\newcommand{\ucne}{Department of Nuclear Engineering, University of California, Berkeley, CA, USA}

\newcommand{\duke}{Department of Physics, Duke University, Durham, NC, USA}
\newcommand{\ncsu}{Department of Physics, North Carolina State University, Raleigh, NC, USA}
\newcommand{\ornl}{Oak Ridge National Laboratory, Oak Ridge, TN, USA}
\newcommand{\ou}{Research Center for Nuclear Physics and Dept. of Physics, Osaka University, Ibaraki, Osaka, Japan}
\newcommand{\pnnl}{Pacific Northwest National Laboratory, Richland, WA, USA}
\newcommand{\sdsmt}{South Dakota School of Mines and Technology, Rapid City, SD, USA}
\newcommand{\usc}{Department of Physics and Astronomy, University of South Carolina, Columbia, SC, USA}
\newcommand{\usd}{Department of Earth Science and Physics, University of South Dakota, Vermillion, SD, USA}
\newcommand{\ut}{Department of Physics and Astronomy, University of Tennessee, Knoxville, TN, USA}
\newcommand{\tunl}{Triangle Universities Nuclear Laboratory, Durham, NC, USA}

\author{
C.E.~Aalseth\address[pnnl]{\pnnl},
E.~Aguayo\addressmark[pnnl],
M.~Amman\address[lbnl]{\lbnl},
F.T.~Avignone~III\address[usc]{\usc}\address[ornl]{\ornl},
H.O.~Back\address[ncsu]{\ncsu}\address[tunl]{\tunl},
X. Bai\address[sdsmt]{\sdsmt},  
A.S.~Barabash\address[ITEP]{\ITEP},
P.S.~Barbeau\address[uchic]{\uchic},
M.~Bergevin\addressmark[lbnl],
F.E.~Bertrand\addressmark[ornl],
M.~Boswell\address[lanl]{\lanl}, 
V.~Brudanin\address[JINR]{\JINR},
W.~Bugg\address[ut]{\ut},
T.H.~Burritt\address[uw]{\uw},
M.~Busch\address[duke]{\duke}\addressmark[tunl],	
G.~Capps\addressmark[ornl],		
Y-D.~Chan\addressmark[lbnl],
J.I.~Collar\addressmark[uchic],
R.J.~Cooper\addressmark[ornl],
R.~Creswick\addressmark[usc],
J.A.~Detwiler\addressmark[lbnl],
J. Diaz\addressmark[uw],	
P.J.~Doe\addressmark[uw],
Yu.~Efremenko\addressmark[ut],
V.~Egorov\addressmark[JINR],
H.~Ejiri\address[ou]{\ou},
S.R.~Elliott\addressmark[lanl],
J.~Ely\addressmark[pnnl],
J.~Esterline\addressmark[duke]\addressmark[tunl],
H.~Farach\addressmark[usc],
J.E.~Fast\addressmark[pnnl],
N.~Fields\addressmark[uchic], 
P.~Finnerty\address[unc]{\unc}\addressmark[tunl],
F.M.~Fraenkle\addressmark[unc]\addressmark[tunl], 
V.M.~Gehman\addressmark[lanl],
G.K.~Giovanetti\addressmark[unc]\addressmark[tunl],  
M.~Green\addressmark[unc]\addressmark[tunl],  
V.E.~Guiseppe\address[usd]{\usd}\thanks{Corresponding author: vincente.guiseppe@usd.edu}\thanks{The author acknowledges the support of the U.S. Department of Energy under Award Number DE-SCOO05054 and the University of South Dakota.},	
K.~Gusey\addressmark[JINR],
A.L.~Hallin\address[alberta]{\alberta},
G.C. Harper\addressmark[uw],	
R.~Hazama\addressmark[ou],
R.~Henning\addressmark[unc]\addressmark[tunl],
A.~Hime\addressmark[lanl],
H. Hong\addressmark[sdsmt],  
E.W.~Hoppe\addressmark[pnnl],
T.W.~Hossbach\addressmark[pnnl], 
S. Howard\addressmark[sdsmt],  
M.A.~Howe\addressmark[unc]\addressmark[tunl],
R.A.~Johnson\addressmark[uw],
K.J.~Keeter\address[blhill]{\blhill},
M.~Keillor\addressmark[pnnl],
C.~Keller\addressmark[usd],
J.D.~Kephart\addressmark[pnnl], 
M.F.~Kidd\addressmark[lanl],	
A. Knecht\addressmark[uw],	
O.~Kochetov\addressmark[JINR],
S.I.~Konovalov\addressmark[ITEP],
R.T.~Kouzes\addressmark[pnnl],
B.H.~LaRoque\addressmark[lanl],	
L.~Leviner\addressmark[ncsu]\addressmark[tunl],
J.C.~Loach\addressmark[lbnl],	
P.N.~Luke\addressmark[lbnl],
S.~MacMullin\addressmark[unc]\addressmark[tunl],
M.G.~Marino\addressmark[uw],
R.D.~Martin\addressmark[lbnl],	
D. Medlin\addressmark[sdsmt],  
D.-M.~Mei\addressmark[usd],
H.S.~Miley\addressmark[pnnl],
M.L.~Miller\addressmark[uw], 
L.~Mizouni\addressmark[usc]\addressmark[pnnl],  
A.W.~Myers\addressmark[pnnl],
M.~Nomachi\addressmark[ou],
J.L.~Orrell\addressmark[pnnl],
D. Peterson\addressmark[uw],	
D.G.~Phillips II\addressmark[unc]\addressmark[tunl],  
A.W.P.~Poon\addressmark[lbnl],
O. Perevozchikov\addressmark[usd],	
G. Perumpilly\addressmark[usd],   
G.~Prior\addressmark[lbnl],
D.C.~Radford\addressmark[ornl],
D.~Reid\addressmark[pnnl],	
K.~Rielage\addressmark[lanl],
R.G.H.~Robertson\addressmark[uw],
L.~Rodriguez\addressmark[lanl],
M.C.~Ronquest\addressmark[lanl],	
H.~Salazar\addressmark[lanl],	
A.G.~Schubert\addressmark[uw],
T.~Shima\addressmark[ou],
M.~Shirchenko\addressmark[JINR],
V. Sobolev\addressmark[sdsmt],  
D.~Steele\addressmark[lanl],	
J.~Strain\addressmark[unc]\addressmark[tunl],
G.~Swift\addressmark[duke]\addressmark[tunl],	
K.~Thomas\addressmark[usd],		
V.~Timkin\addressmark[JINR],
W.~Tornow\addressmark[duke]\addressmark[tunl],
T.D.~Van Wechel\addressmark[uw],
I.~Vanyushin\addressmark[ITEP],
R.L.~Varner\addressmark[ornl],  
K.~Vetter\address[ucne]{\ucne}\addressmark[lbnl],
K.~Vorren\addressmark[unc]\addressmark[tunl], 
J.F.~Wilkerson\addressmark[unc]\addressmark[tunl]\addressmark[ornl],    
B.A. Wolfe\addressmark[uw],	
W. Xiang\addressmark[usd],		
E.~Yakushev\addressmark[JINR],
H.~Yaver\addressmark[lbnl],	
A.R.~Young\addressmark[ncsu]\addressmark[tunl],
C.-H.~Yu\addressmark[ornl],
V.~Yumatov\addressmark[ITEP] and
C.~Zhang\addressmark[usd]\\
}

\begin{document}

\begin{abstract}
The M{\sc
ajorana} Collaboration is assembling an array of HPGe
detectors to search for neutrinoless double-beta decay in
$^{76}$Ge. 
Initially, M{\sc ajorana} aims to construct a
prototype module to demonstrate
the potential of a future 1-tonne experiment. The design and
potential reach of this prototype \DEM\ module are presented. \vspace{1pc}
\end{abstract}

\maketitle

\section{Introduction}
The \MJ\ Collaboration is fielding a neutrinoless double-beta decay (\nonubb) experiment using the well-established technique
of searching for \nonubb\ decay in high-purity
Ge-diode radiation detectors that play both roles of source and
detector. The technique maximizes the source to total mass ratio and benefits from excellent energy resolution (0.16\% at 2.039 MeV). Ge detectors can be enriched in the \bb-decay isotope $^{76}$Ge from 7.44\% to more than 86\%.
Ge-based \nonubb\ experiments have established the best half-life limits and the most restrictive constraints on the effective Majorana neutrino mass \cite{aal02a,bau99}. 

The \MJ\ collaboration is currently pursuing R\&D aimed at a tonne scale,
$^{76}$Ge \nonubb-decay experiment. The goal of a tonne-scale experiment would be to probe the effective neutrino mass in the inverted hierarchy of 20-40 meV. A discovery of  \nonubb\ would indicate the Majorana nature of neutrinos and establish lepton number violation.  The desired mass scale corresponds to a decay half-life of 10$^{27}$ yr and will require backgrounds below 1 count/tonne/yr in the 4-keV region of interest. We are currently cooperating with the GERDA  Collaboration \cite{sch05} with the aim to jointly prepare for a single international tonne-scale Ge-based experiment utilizing the best technologies of the two collaborations. The GERDA approach uses a large cryostat with liquid cryogen shielding.

For the R\&D phase, the \MJ\ collaboration is constructing
a \DEM\ module of $^{76}$Ge crystals contained
in an ultra-low-background structure deep underground.  
The  \DEM\ reference design is modular with two 
cryostats containing a combination of enriched (86\%-enrichment) and unenriched
(natural) Ge crystals. The detectors will
be mounted in a string-like arrangement in ultra-pure vacuum cryostats made from  electro-formed copper.  The cryostats will be enclosed in a graded, multilayer Cu and Pb passive shield and an active muon veto to eliminate external backgrounds.
By using up to 30 kg of enriched $^{76}$Ge crystals, the \nonubb\ claim \cite{kla06} can be tested within 2 yr of running and the background requirement of a tonne-scale experiment can be demonstrated.



\section{Backgrounds}
\label{sect:bg}
Mitigation of backgrounds is crucial to the success of any rare
decay search. For the case of Ge solid-state
detectors, decades of research and recent advances have yielded a host of techniques to
reduce backgrounds. 
Pulse shape discrimination and detector granularity techniques make use of the different
spatial distributions of energy deposition between
double-beta decay events and most background signals. Double-beta decay energy deposition occurs within a
small volume ($<$1~mm$^{3}$) and  hence is single site.
Background signals arising from radioactive decays may include 
one or more $\gamma$ rays, which frequently undergo multiple scatters on the centimeter
scale and are multiple site interactions. Therefore, analysis techniques that can identify multiple-site interactions can reject most backgrounds.
In addition, backgrounds are suppressed through the use of ultra-pure
materials for the construction of detector components in the
proximity of the crystals, and shielding the detectors from external
natural and cosmogenic sources.
 Monte Carlo radiation transport simulations \cite{bos10} are used to build a background model based on achievable material purities and guide the design of the experiment.

The isotopes produced by cosmic activation are particularly problematic. At the Earth's surface, cosmic rays can activate isotopes in detector materials. These backgrounds will be reduced by carefully limiting exposure of detectors and susceptible shielding materials above ground. 
Production of detector components will take place underground when possible.
Once underground, backgrounds can be produced from the remaining muon-induced hard, secondary neutrons  interacting in the detector and shielding materials.  For example, $(n,n'\gamma)$ reactions will become important for tonne-scale \nonubb\ decay experiments depending on the depth of the experiment \cite{mei06}. Previously unknown cross sections of $(n,n'\gamma)$ and reactions on Pb shielding important for a \nuc{76}{Ge} \nonubb\ decay experiment have been measured \cite{gui09} and will be included in background simulations.

\section{Detectors}
\label{sec:det}
The \MJ\ \DEM\ will use modified Broad Energy Ge (BEGe) detectors \cite{canberra}, which are a variation on the 
p-type, point-contact (\ppc) detector design \cite{luk89,bar07}.
\ppc\ detectors offer excellent interaction site multiplicity information, superior energy
resolution and low energy threshold. 
A \ppc\ detector is a right circular cylindrical Ge crystal with a point-contact in
the center of the detector face resulting in a low capacitance and 
electronic noise. It has been shown that the low noise results in a low energy
trigger threshold and excellent energy resolution  making the detector sensitive to WIMP-nucleon elastic scattering \cite{aal08}.
The \ppc\ detector design has a low electric field throughout the bulk
of the crystal and a weighting potential that is sharply peaked
near the point contact. The resulting long drift  times and fast electric signal 
of the charge cloud at the central electrode provides a very effective active 
background separation of multiple interactions by pulse-shape analysis \cite{bar07,bud09,coo10}.

Compared to other detector technologies that offer a similar pulse shape analysis capability, \ppc-style detectors are cost-effective, have increased 
manufacturing speed, a simpler contact  scheme, and a lower energy threshold.
We have succeeded in validating field calculation codes and
simulations that allow us to determine the impurity gradient
requirements and the dimensions to which such detectors can reliably
be manufactured. In addition to a variety of \ppc\ prototypes across the collaboration, we have acquired 20 kg of natural Ge modified BEGe detectors to populate the initial cryostat of the \DEM. The detectors measure 70 mm in diameter and 30 mm in height and define the primary detector configuration.

\section{Summary}
Our proposed method uses the well-established technique of searching for neutrinoless double-beta decay in high purity
Ge-diode radiation detectors that play both roles of source and detector.
The use of \ppc\ Ge detectors present advances in background rejection and a significantly lower energy threshold than conventional Ge detector technologies. The lower energy threshold opens up a broader and exciting physics program including searches for dark matter and axions concurrent with the double-beta decay search.
The \DEM\ should establish that the backgrounds are low enough to justify scaling to tonne-scale experiment, probe the neutrino effective mass region above 100 meV, and search the low energy region with a sensitivity to dark matter. The \DEM\ will be sited at the 4850-ft level (4200 m.w.e) of the Sanford Underground Laboratory at Homestake and preparations for construction are currently underway.

\bibliographystyle{elsart-num}
\bibliography{mymj_now2010}

\end{document}